\documentclass[aoas,preprint]{imsart}

\RequirePackage{amsthm,amsmath,amsfonts,amssymb}
\RequirePackage[authoryear]{natbib}
\RequirePackage[colorlinks,citecolor=blue,urlcolor=blue]{hyperref}
\RequirePackage{graphicx}

\startlocaldefs

\usepackage{mathtools}

\theoremstyle{remark}

\newcommand{\M}[1]{\boldsymbol{#1}}  
\newcommand{\V}[1]{\boldsymbol{#1}}  
\newcommand{\X}[1]{\mathbf{#1}}  
\newcommand{\Unif}[0]{\textrm{Uniform}}

\newcommand{\dataset}{{\cal D}}
\newcommand{\EX}[0]{\mathbb{E}}

\DeclareMathOperator*{\VA}{\mathbb{V}ar}

\allowdisplaybreaks[1]

\allowdisplaybreaks
\endlocaldefs

\begin{document}

\begin{frontmatter}
\title{Gaussian process surrogate modeling with manipulating factors for carbon nanotube growth experiments }
\runtitle{GP surrogate for carbon nanotube growth experiments }

\begin{aug}
\author[A]{\fnms{Chiwoo} \snm{Park}\ead[label=e1, mark]{cpark5@fsu.edu}}
,
\author[C]{\fnms{Rahul} \snm{Rao}\ead[label=e3]{rahul.rao.1.ctr@us.af.mil}}
,
\author[B]{\fnms{Pavel} \snm{Nikolaev}\ead[label=e3]{}}
\and
\author[C]{\fnms{Benji} \snm{Maruyama}\ead[label=e3]{benji.maruyama@us.af.mil}}

\address[A]{Department of Industrial and Manufacturing Engineering,
Florida State University,
\printead{e1}}

\address[B]{Cornerstone Research Group
}

\address[C]{Materials and Manufacturing Directorate,  
	Air Force Research Lab,
}
\end{aug}

\begin{abstract}
Running physical experiments is often expensive in time and cost, for which physical experiments can be replaced with the prediction from a surrogate model estimated using a few past experimental outcomes. This paper presents a new Gaussian process (GP) surrogate modeling for predicting the outcome of a physical experiment where some experimental inputs are controlled by other manipulating factors. Particularly, we are interested in the case where the control precision is not very high, so the input factor values vary significantly even under the same setting of the corresponding manipulating factors. The case is observed in our main application to carbon nanotube growth experiments, where one experimental input among many is manipulated by another manipulating factors, and the relation between the input and the manipulating factors significantly varies in the dates and times of operations. Due to this variation, the standard GP surrogate that directly relates the manipulating factors to the experimental outcome does not provide a great predictive power on the outcome. At the same time, the GP model relating the main factors to the outcome directly is not appropriate for the prediction purpose because the main factors cannot be accurately set as planned for a future experiment. Motivated by the carbon nanotube example, we propose a two-tiered GP model, where the bottom tier relates the manipulating factors to the corresponding main factors with potential biases and variation independent of the manipulating factors, and the top tier relates the main factors to the experimental outcome. Our two-tier model explicitly models the propagation of the control uncertainty to the experimental outcome through the two GP modeling tiers. We present the inference and hyper-parameter estimation of the proposed model. The proposed approach is illustrated with the motivating example of a closed-loop autonomous research system for carbon nanotube growth experiments, and the test results are reported with the comparison to a benchmark method, i.e. a standard GP model. 
\end{abstract}

\begin{keyword}
\kwd{Surrogate Modeling}
\kwd{Input Uncertainty}
\kwd{Control Uncertainty}
\kwd{Gaussian Process}
\end{keyword}

\end{frontmatter}

\section{Introduction} \label{sec:intro}
Carbon nanotubes are promising for many practical applications due to their superior mechanical, electrical and magnetic properties \citep{rao2018carbon}. A large scale chemical growth of carbon nanotubes has been of great interest due to the large practical needs, but it is mainly limited by the difficulty of producing them with controlled structures and properties. We are mainly interested in creating a surrogate model of a carbon nanotube growth experiment that relates experimental conditions to the growth outcome. Once created, the surrogate model can provide the prediction of the experimental outcome for a given experimental condition, replacing expensive physical experiments. The surrogate model can be potentially combined with a Bayesian optimization in a research robot, Autonomous Research System (ARES) that performs closed-loop carbon nanotube (CNT) growth experiments, to search for the better experimental conditions that induce more carbon nanotube growths in a closed loop fashion. 

In this paper, we use a Gaussian process (GP) surrogate modeling approach, which has been a popular nonparametric approach for surrogate modeling of physical or computer experiments \citep{barton1998simulation, ankenman2010stochastic} as well as nonlinear regression \citep{rasmussen2006gaussian, park2018patchwork} and global optimization \citep{Jones1998}. In the surrogate model, the response variable represents the experimental outcome that quantifies the carbon nanotube growth, and the explanatory variables prescribe the experimental factors of interest that affect the experimental outcome. One peculiarity here is that some experimental conditions are indirectly manipulated by other factors. For example, the reaction temperature in a carbon nanotube growth process is controlled by applying a heat source such as a laser irradiation, and the applied laser power is tuned to achieve the desirable temperature. In this example, the temperature is a main factor presumably related to the experiment outcome, and the laser power is the factor treated to manipulate the main factor, which will be referred to as a manipulating factor in this paper. When a surrogate model is intended for predicting a future experimental outcome or optimizing it, surrogate modeling with manipulating factors is more useful than that with main factors, because the main factors cannot be precisely set as planned. 

A major challenge in surrogate modeling with manipulating factors is that the manipulation precision to main factors is often limited. The low precision could bring up another source of variation in the surrogate modeling. The precision may vary from day-to-day, creating mean shifts and variation in the main factor values even under the exactly same setting of the corresponding manipulating factor. This would result in a larger variation in the response variable. With this larger variation, applying conventional GP surrogate modeling directly with manipulating factors could lead to a bad representation of the experiment. This paper presents a new approach to a GP surrogate modeling with manipulating factors. 

The major development in this paper is related to the existing works regarding the GP modeling with input uncertainty because the lack of the control precision on experimental inputs would create uncertainty on the input factors. However, what has been discussed in the literature is a lot simpler than what we are going to study. In the literature, the GP modeling with input uncertainty has been mainly discussed in the context of noisy inputs \citep{mchutchon2011gaussian, damianou2016variational}, where an input $x$ is identifiable from its observed value ${u}$ up to a random noise variation ${\delta}$, 
\begin{equation} \label{eq:inputunc}
{x} = {u} + {\delta}.
\end{equation}
In this case, the input uncertainty given the observation is quantified by a normal distribution on $x$, $\mathcal{N}(x; {u}, \sigma^2)$. A more general multivariate normal case was discussed in \citet{girard2003gaussian}. In our problem to the GP modeling with manipulating factors, the main factor ${x}$ manipulated by $u$ is modeled as a function of the manipulating factor ${u}$, but it is also affected by a source of variation due to exogenous variables $\omega$, 
\begin{equation} \label{eq:covariates}
{x} = g({u}) + {\delta}(\omega),
\end{equation}
where $g(\cdot)$ is an unknown effect of the manipulating factor to the main factor, and $\delta(\cdot)$ is the variation due to the exogenous variables $\omega$; the $\delta$ is not purely white Gaussian noise but correlated noises, which are modeled in this paper as a Gaussian process. The existing model \eqref{eq:inputunc} can be seen as a special case of \eqref{eq:covariates} when $g$ is a simple linear function and $\delta$ is an independent Gaussian noise. Along with the Gaussian process surrogate model relating the main factors to the response variable, the GP model \eqref{eq:covariates} for some input variables forms a two-tiered GP model, where the bottom tier relates the manipulating factors to the corresponding main factors, and the top tier relates the main factors to the response variable. The two-tier model explicitly models the propagation of the control uncertainty to the experimental outcome through the two GP modeling tiers. 

Another existing researches relevant to this paper is the linked Gaussian process model \citep{kyzyurova2018coupling}, where an input of a Gaussian process regression model is modeled as another Gaussian process model. This is similar to what we are studying. If we find the correspondence of the existing work to the proposed model \eqref{eq:covariates}, it would be
\begin{equation*}
x = \tilde{\delta}(u),
\end{equation*}
where $\tilde{\delta}(\cdot)$ is a Gaussian process with manipulating factors $u$. Therefore, the model is not same as what we are trying to study for our motivating example. In addition, the existing work mainly provided the analytical closed form solution when a squared exponential covariance function is used. Since an exponential covariance function works better for our motivating example, we also develop the similar results with an exponential covariance function or equivalently the Matérn covariance function with $\nu = 0.5$. In addition to the existing work, linked models in some form or another have been studied in an area of engineering called partitioned analysis. \citet{stevens2016experiment} studied validation and uncertainty quantification for weakly coupled systems, but without surrogate modeling, and \citet{stevens2018statistical} performed similar works in the context of model calibration when the linked models have GP surrogates.

The remainder of this paper is organized as follows. Section \ref{sec:method} describes our modeling approach and model estimation. Section \ref{sec:numerical} presents numerical examples to evaluate the proposed approach with simulated datasets and a real dataset. Section \ref{sec:conclusion} concludes this paper with discussion.

\section{General Approach} \label{sec:method}
Consider an experiment that involves $d$ main factors and a real response variable $y$. For the time being, we assume only one of the main factors is manipulated by a $p$-dimensional manipulating factor $u$ for a concise description of the proposed approach, and its extension for a more general case with multiple factors manipulated will be discussed in Section \ref{sec:ext}. Let $z$ denote the main factor manipulated by $u$ and let $x$ denote the other main factors. Here we used regular fonts for multivariate quantities, $x$ and $u$, and bold fonts will be reserved to denote a collection of these quantities later. The control precision of the main factor $z$ is limited in that $z$ is not only controlled by $u$ but also influenced by uncontrollable (but still observable) exogenous variables $\omega$,
\begin{equation} \label{eq:z}
z = g(u) + \delta(\omega),
\end{equation}
where $g$ is an unknown function that relates $u$ to $z$, and $\delta$ is independent of the manipulating factor $u$ but depends on the exogenous variable $\omega$. Depending on $\omega$, $z$ is shifted by a random quantity $\delta(\omega)$ that imposes the mean shift in $z$ and the variation of $z$. An example of $\omega$ is dates of control operations or different batches of experiment operations. For different dates, the resulting $z$ for an applied control $u$ can be different. 

We like to estimate an unknown metamodel that relates $x$ and $u$ to $y$ with data from $N$ past experiments, $\dataset = \{(x_i, u_i, \omega_i, z_i, y_i), i = 1,\ldots, N\}$. Here we assume that the manipulated main factor $z$ is still measurable during experiments, e.g., temperature is measured for a given laser power. Since $z$ is measured during experiments, its values are only available for the past experiments but are not available for any future experiments being performed yet. Therefore, $z$ cannot be used as a predictor to predict future experiments. To sum up, $x, u$ and $\omega$ are set before experiments, while $z$ and $y$ are measured during experiments.  Here we used regular fonts for multivariate quantities, $x$ and $u$, and reserved bold fonts for the collections of the factor values from past experiments, e.g. $\X{x} = [x_i, i = 1,\ldots, N]$, $\X{\tilde{x}} = [(x_i, z_i), i = 1,\ldots, N]$, $\X{z} = [z_i, i = 1,\ldots, N]$, $\X{u} = [u_i, i = 1,\ldots, N]$, and $\V{\omega} = [\omega_i, i = 1,\ldots, N]$.

Let $f$ denote an unknown function that relates $x$ and $z$ to $y$. The observed response variable $y_i$ is assumed a noisy observation of $f$,
\begin{equation*}
y_i = f(x_i, z_i) + \epsilon_i \mbox{ for } i = 1,...,N, 
\end{equation*}
where $\epsilon_i \sim \mathcal{N}(0, \sigma^2)$. We assume that the unknown function $f$ is a realization of a zero-mean Gaussian process with covariance function $c(x, x')$. The observed factor $z_i$ is the observation of the assumed model \eqref{eq:z} at $u_i$ with random variation $\delta(\omega_i)$, 
\begin{equation*}
z_i = g(u_i) + \delta(\omega_i),\mbox{ for } i = 1,...,N.
\end{equation*}
Unless other noisy factors involve, the relation of manipulating factors $u$ to $z$ is deterministic. We use a general semi-parametric model to represent the deterministic relation,
\begin{equation} \label{eq:param}
\begin{split}
g(u) = \sum_{q=1}^Q \beta_q \pi_q(u), 
\end{split}
\end{equation} 
where $\{\pi_q(u)\}$ are the set of the fixed basis functions to represent the linear spaces that $g(u)$ belongs to, and the parameter set, $\{\beta_q\}$, are unknown coefficients to represent the two functions in the linear space. The random variation $\delta$ due to noise factors $\omega$ is modeled as a Gaussian process with mean function $\rho(\omega)$ and covariance function $k(\omega, \omega')$. Accordingly, $z_i$ is a realization of a GP with mean function $g(u) + \rho(\omega)$ and covariance function $k(\omega, \omega')$. We use a constant mean function for $\delta(\omega)$, 
\begin{equation*}
\rho(\omega) = \alpha,
\end{equation*}
because taking a general nonparametric or semi-parametric form for $\rho$ would cause the well identifiability issue in determining $\rho$. Given observations of $z_i$, identifying $g$ and $\delta$ is quite feasible, mainly because the two terms depend on different independent variables. We also present numerically in Section \ref{sec:sim_first} as for how $g$ and $\delta$ are separately estimable. 

Given data $\dataset$, we like to predict the value of $f$ when $x$ and $u$ are specified to $x_*$ and $u_*$, which we denote by $f_*$. To be more specific, we want the posterior predictive distribution of $f_*$ given $\dataset$. Let $p(f_* | x_*, u_*, \dataset)$ denote the density of the posterior predictive distribution. It can be achieved by the following integration,
\begin{equation} \label{eq:pos}
p(f_* | x_*, u_*, \omega_*, \dataset) = \int p(f_* | x_*, z_*, \dataset) p(z_* | u_*, \omega_*, \dataset) dz_*,
\end{equation}
where $z_*$ denotes the unknown value of $z$ given $u=u_*$ and $\omega=\omega_*$, $p(z_* | u_*, \omega_*, \dataset)$ denotes the posterior density of $z_*$ given $\dataset$, and $p(f_* | x_*, z_*, \dataset)$ is the conditional density of $f_*$ given $z_*$ and $\dataset$. 

The two parts of the integrand can be obtained using the standard formula for GP regression. Denote $\V{y} = [y_1,y_2,\dots,y_N]^T$.  With the GP prior on $f$, the joint distribution of $(f_*, \V{y})$ is 
\begin{equation} \label{eq:joint_y}
f_*, \V{y} | x_*, z_*, \X{\tilde{x}} \sim
\mathcal{N}\left( \left[ \begin{array} {c} 
0 \\
\V{0} \end{array} \right], \left[ \begin{array} {c c} c_{**}
& \V{c}_{\X{\tilde{x}}*}^T \\ \V{c}_{\X{\tilde{x}}*}
& \sigma^2\M{I} + \M{C}_{\X{\tilde{x}\tilde{x}}} \end{array} \right] \right),
\end{equation}
where $c_{**} = c((x_*,z_*), (x_*,z_*))$, $\V{c}_{\X{\tilde{x}}*} = [c((x_1,z_1),(x_*, z_*)), \dots, c((x_N,z_N), (x_*,z_*))]^T$ and $\M{C}_{\X{\tilde{x}\tilde{x}}}$ is an $N \times N$ matrix with $(i,j)^{th}$ entry $c((x_i,z_i), (x_j,z_j))$. Applying the Gaussian conditioning formula to the joint distribution gives the posterior distribution of $f_*$,
\begin{equation}\label{eq:pred-dist}
f_* |x_*, z_*, \dataset \sim \mathcal{N}(
\mu(x_*, z_*), s^2(x_*, z_*)).
\end{equation}
where
\begin{equation*}
\begin{split}
& \mu(x_*, z_*) = \V{c}_{\X{\tilde{x}}*}^T (\sigma^2\M{I} + \M{C}_{\X{\tilde{x}\tilde{x}}})^{-1} \V{y} \quad \mbox{ and } \\
& s^2(x_*, z_*) = c_{**} - \V{c}_{\X{\tilde{x}}*}^T (\sigma^2\M{I} + \M{C}_{\X{\tilde{x}\tilde{x}}})^{-1}
\V{c}_{\X{\tilde{x}}*}. 
\end{split}
\end{equation*}
Similarly, Given $u = u_*$ and $\omega = \omega_*$, we have the joint distribution of $(z_*, \V{z})$, 
\begin{equation}  \label{eq:joint_z}
z_*, \V{z} | u_*, \omega_*, \X{u}, \V{\omega} \sim
\mathcal{N}\left( \left[ \begin{array} {c} 
\rho(\omega_*) + g(u_*) \\
\V{\rho} + \V{g} \end{array} \right], \left[ \begin{array} {c c} 
k_{**} & \V{k}_{\V{\omega}*}^T \\ 
\V{k}_{\V{\omega}*} & \M{K}_{\V{\omega\omega}} \end{array} \right] \right),
\end{equation}
where $\V\rho = [\rho(\omega_1), \ldots, \rho(\omega_N)]^T$, $\V{g} = [g(u_1), \ldots, g(u_N)]^T$, $\V{k}_{\V{\omega}*} = (k(\omega_1,\omega_*), \dots, k(\omega_N, \omega_*))^T$, $k_{**} = k(\omega_*, \omega_*)$,  and $\M{K}_{\V{\omega\omega}}$ is an $N \times N$ matrix with $(i,j)^{th}$ entry $k(\omega_i, \omega_j)$. The posterior distribution of $z_*$ is
\begin{equation}\label{eq:pred-dist2}
z_* |u_*, \omega_*, \dataset \sim 
\mathcal{N}(
\nu_*,
t^2_*),
\end{equation}
where $\nu_* = \rho(\omega_*) + g(u_*) + \V{k}_{\V{\omega}*}^T \M{K}_{\V{\omega\omega}}^{-1} (\X{z} - \V\rho - \V{g} )$ and $t^2_* = k_{**} - \V{k}_{\V{\omega}*}^T \M{K}_{\V{\omega\omega}}^{-1}\V{k}_{\V{\omega}*}$. 

Given these, the integral \eqref{eq:pos} for the posterior predictive distribution cannot be analytically solved, but its posterior mean and variance can be evaluated using the properties of the conditional expectation, 
\begin{equation} \label{eq:posexp}
\begin{split}
\EX[f_*|x_*, u_*, \omega_*, \dataset] =& \EX[\EX[f_*|x_*,z_*, \dataset]|u_*, \omega_*, \dataset] \\
=& \EX[\mu(x_*, z_*) |u_*, \omega_*, \dataset], \\
\VA[f_*|x_*, u_*, \omega_*, \dataset] =& \EX[\VA[f_*|x_*,z_*, \dataset]|u_*, \omega_*, \dataset] \\
& + \VA[\EX[f_*|x_*,z_*, \dataset]|u_*, \omega_*, \dataset] \\
=& \EX[s^2(x_*, z_*) |u_*, \omega_*, \dataset] \\
& + \EX[\mu(x_*, z_*)^2|u_*, \omega_*, \dataset] - \EX^2[\mu(x_*, z_*)|u_*, \omega_*, \dataset].
\end{split}
\end{equation}
The posterior predictive distribution of $f_*$ is approximated locally at $x_*$ and $u_*$ by a multivariate normal distribution with mean and variance given in \eqref{eq:posexp}. We can derive the analytical forms of the posterior mean and covariance when the covariance function $c$ is a squared exponential covariance or exponential covariance function. Otherwise, they should be numerically approximated by Monte Carlo integration. In the two subsequent subsections, we will derive the analytical forms for two popular covariance functions, an exponential covariance function in Section \ref{sec:exp} and a squared exponential covariance function in Section \ref{sec:exp2}. We will discuss the maximum likelihood estimates of the model parameters in Section \ref{sec:hyper}. Section \ref{sec:ext} will include how the proposed model can be extended for a general case of multiple main factors manipulated by other factors. 

\subsection{Case for exponential covariance function $c$} \label{sec:exp}
Suppose that the covariance function $c$ is an anisotropic exponential covariance function as described below, 
\begin{equation} \label{eq:cov_c}
\begin{split}
c((x_i,z_i), (x_j, z_j)) & = c_x(x_i, x_j) c_z(z_i, z_j), \mbox{ where } \\
& c_x(x_i, x_j) = \exp\left\{ - \sum_{l=1}^{d-1} \frac{|x_{il}-x_{jl}|}{b_l} \right \}, \\ 
& c_z(z_i, z_j) = \exp\left\{ - \frac{|z_{i}-z_{j}|}{b_z} \right \}, \mbox{ and }
\end{split}
\end{equation}
$\{b_l: l = 1,\ldots, d-1\}$ and $b_z$ are the hyperparameters of the covariance function. 
Let $\V{c}_{\X{x}*} = [c_x(x_1,x_*), \dots, c_x(x_N,x_*)]^T$ and $\V{c}_{\X{z}*} = [c_z(z_1,z_*), \dots, c_z(z_N,z_*)]^T$. Based on \eqref{eq:posexp}, the predictive mean of $f_*$ is 
\begin{equation*}
\begin{split}
\EX[f_*|x_*, u_*, \omega_*, \dataset] &= \EX[\mu(x_*, z_*) |u_*, \omega_*, \dataset] \\
&= \EX[\V{y}^T (\sigma^2\M{I} + \M{C}_{\X{\tilde{x}\tilde{x}}})^{-1} \V{c}_{\X{\tilde{x}}*}] \\
& = \V{y}^T (\sigma^2\M{I} + \M{C}_{\X{\tilde{x}\tilde{x}}})^{-1} \int (\V{c}_{\X{x}*} \odot \V{c}_{\X{z}*}) p(z_* |u_*, \omega_*, \dataset) dz_* \\
& = \V{y}^T (\sigma^2\M{I} + \M{C}_{\X{\tilde{x}\tilde{x}}})^{-1} \left(\V{c}_{\X{x}*} \odot \int \V{c}_{\X{z}*}p(z_* |u_*, \omega_*, \dataset) dz_* \right),
\end{split}       
\end{equation*}
where $\odot$ is the element-wise matrix product operator. To simplify the expression, we define new symbols. Let $\tilde{\V{y}} = (\sigma^2\M{I} + \M{C}_{\X{\tilde{x}\tilde{x}}})^{-1} \V{y}$, and $\V{\tilde{c}}_{\X{z}*}$ is a $N \times 1$ vector with its $i$th element equal to $\int c_z(z_i, z_*) p(z_* |u_*, \omega_*, \dataset) dz_*$. With the new symbols, the predictive mean is 
\begin{equation*}
\begin{split}
\EX[\mu(x_*, z_*) |u_*, \omega_*, \dataset] = \V{\tilde{y}}^T (\V{c}_{\X{x}*} \odot \V{\tilde{c}}_{\X{z}*}).
\end{split}       
\end{equation*}
The $i$th element of the vector $\V{\tilde{c}}_{\X{z}*}$ can be analytically evaluated as
\begin{align*}
(\V{\tilde{c}}_{\X{z}*})_i &= \int c_z(z_i, z_*) p(z_* |u_*, \omega_*, \dataset) dz_* \\
&= \int_{z_* \le z_i} \exp\left\{ - \frac{z_{i}-z_{*}}{b_z} \right \}p(z_* |u_*, \omega_*, \dataset) dz_* \\
& \qquad + \int_{z_* > z_i} \exp\left\{  \frac{z_{i}-z_{*}}{b_z} \right \}p(z_* |u_*, \omega_*, \dataset) dz_* \\
&= \int_{z_* \le z_i} \exp\left\{ - \frac{z_{i}-z_{*}}{b_z} \right \} \frac{1}{\sqrt{2\pi t^2_*}} \exp\left\{ - \frac{ (z_{*}-\nu_*)^2}{2 t^2_*} \right \} dz_* \\
& \qquad + \int_{z_* > z_i} \exp\left\{ \frac{z_{i}-z_{*}}{b_z} \right \} \frac{1}{\sqrt{2\pi t^2_*}} \exp\left\{ - \frac{ (z_{*}-\nu_*)^2}{2 t^2_*} \right \} dz_* \\
&= \exp\left\{ \frac{t_*^2 + 2(v_* - z_i)b_z}{2b_z^2}  \right\} \Phi\left(\frac{z_i-v_*}{t_*} - \frac{t_*}{b_z}\right) \\
&\qquad +\exp\left\{ \frac{t_*^2 - 2(v_* - z_i)b_z}{2b_z^2}  \right\} \left[ 1 - \Phi\left(\frac{z_i-v_*}{t_*} + \frac{t_*}{b_z}\right) \right].
\end{align*}
Likewise, the predictive variance at $x_*$ and $u_*$ is 
\begin{equation*}
\begin{split}
\VA[f_*|x_*, u_*, \omega_*, \dataset] = & \EX[s^2(x_*, z_*) |u_*, \dataset] + \EX[\mu(x_*, z_*)^2|u_*, \dataset]  - \EX^2[\mu(x_*, z_*)|u_*, \dataset] \\
= & \EX[c_{**} - \V{c}_{\X{\tilde{x}}*}^T (\sigma^2\M{I} + \M{C}_{\X{\tilde{x}\tilde{x}}})^{-1}
\V{c}_{\X{\tilde{x}}*}] + \EX[(\V{y}^T (\sigma^2\M{I} + \M{C}_{\X{\tilde{x}\tilde{x}}})^{-1} \V{c}_{\X{\tilde{x}}*})^2] \\
& - \EX^2[\V{y}^T (\sigma^2\M{I} + \M{C}_{\X{\tilde{x}\tilde{x}}})^{-1} \V{c}_{\X{\tilde{x}}*}] \\
= & 1 + \EX[ - \V{c}_{\X{\tilde{x}}*}^T (\sigma^2\M{I} + \M{C}_{\X{\tilde{x}\tilde{x}}})^{-1}
\V{c}_{\X{\tilde{x}}*}] \\
& + \EX[ \V{c}_{\X{\tilde{x}}*}^T  (\sigma^2\M{I} + \M{C}_{\X{\tilde{x}\tilde{x}}})^{-1} \V{y} \V{y}^T (\sigma^2\M{I} + \M{C}_{\X{\tilde{x}\tilde{x}}})^{-1} \V{c}_{\X{\tilde{x}}*}] \\
& - \EX^2[\V{y}^T (\sigma^2\M{I} + \M{C}_{\X{\tilde{x}\tilde{x}}})^{-1} \V{c}_{\X{\tilde{x}}*}] \\
= & 1 + \int \V{c}_{\X{\tilde{x}}*}^T \M{Q} \V{c}_{\X{\tilde{x}}*} p(z_* |u_*, \omega_*, \dataset) dz_* - \left[\tilde{\V{y}}^T (\V{c}_{\X{x}*} \odot \V{\tilde{c}}_{\X{z}*})\right]^2\\
= & 1 + \mbox{tr}\left[\M{Q} (\V{c}_{\X{x}*} \V{c}_{\X{x}*}^T) \odot \M{\tilde{C}}_{\X{zz}} \right] - \left[\tilde{\V{y}}^T (\V{c}_{\X{x}*} \odot \V{\tilde{c}}_{\X{z}*})\right]^2,
\end{split}       
\end{equation*}
where $\M{Q} = \tilde{\V{y}}\tilde{\V{y}}^T - (\sigma^2\M{I} + \M{C}_{\X{\tilde{x}\tilde{x}}})^{-1}$, and $\M{\tilde{C}}_{\X{zz}}$ is a $N \times N$ matrix with its $(i,j)$th element equal to 
\begin{align*}
\int c_z(z_i, z_*) c_z(z_j, z_*) & p(z_* |u_*, \omega_*, \dataset) dz_* \\
&= \int_{z_* \le z_m} \exp\left\{ - \frac{z_{i}+z_j-2t_{*}}{b_z} \right \}p(z_* |u_*, \omega_*, \dataset) dz_* \\
& \qquad+ \int_{z_m < z_* \le z_M} \exp\left\{ - \frac{z_{M}-z_{m}}{b_z} \right \}p(z_* |u_*, \omega_*, \dataset) dz_* \\
&  \qquad+ \int_{z_* > z_M} \exp\left\{  \frac{z_{i}+z_j-2t_{*}}{b_z} \right \}p(z_* |u_*, \omega_*, \dataset) dz_* \\
&= \exp\left\{ \frac{2t_*^2 + (2t_* - z_i-z_j)b_z}{b_z^2}  \right\} \Phi\left(\frac{z_m-v_*}{t_*} - \frac{2t_*}{b_z}\right) \\
&\qquad + \exp\left\{ - \frac{z_{M}-z_{m}}{b_z} \right \} \left[ \Phi\left(\frac{z_M-v_*}{t_*}\right) - \Phi\left(\frac{z_m-v_*}{t_*}\right) \right] \\
&\qquad +\exp\left\{ \frac{2t_*^2 - (2t_* - z_i-z_j)b_z}{b_z^2}  \right\} \left[ 1 - \Phi\left(\frac{z_M-v_*}{t_*} + \frac{2t_*}{b_z}\right) \right].
\end{align*}
where $z_m = \min\{z_i, z_j\}$ and $z_M = \max\{z_i, z_j\}$.

\subsection{Case for squared exponential covariance function $c$} \label{sec:exp2}
Now consider that the covariance function $c$ is an anisotropic exponential covariance function as described below,
\begin{equation*}
c((x_i,z_i), (x_j, z_j))  = c_x(x_i, x_j) c_z(z_i, z_j), 
\end{equation*}
where 
\begin{equation*}
\begin{split}
& c_x(x_i, x_j) = \exp\left\{ - \sum_{l=1}^{d-1} \frac{(x_{il}-x_{jl})^2}{2b_l^2} \right \} \mbox{ and } \\ 
& c_z(z_i, z_j) = \exp\left\{ - \frac{(z_{i}-z_{j})^2}{2b_z^2} \right \}.
\end{split}
\end{equation*}
Similar to the case with the exponential covariance, the predictive mean at $x_*$ and $u_*$ is in the form of 
\begin{equation*}
\begin{split}
\EX[\mu(x_*, z_*) |u_*, \omega_*, \dataset] = \V{\tilde{y}}^T (\V{c}_{\X{x}*} \odot \V{\tilde{c}}_{\X{z}*}).
\end{split}       
\end{equation*}
where $\V{\tilde{c}}_{\X{z}*}$ is a $N \times 1$ vector with its $i$th element equal to
\begin{equation*}
\begin{split}
(\V{\tilde{c}}_{\X{z}*})_i &= \int c_z(z_i, z_*) p(z_* |u_*, \omega_*, \dataset) dz_* \\
&= \int \exp\left\{ - \frac{(z_{i}-z_{*})^2}{2b_z^2} \right \}p(z_* |u_*, \omega_*, \dataset) dz_* \\
&= (1+t_*^2/b_z^2)^{-1/2} \exp\left\{-\frac{1}{2}\frac{(\nu_x - z_i)^2}{b_z^2 + t_*^2}\right\}.
\end{split}
\end{equation*}
The predictive variance at $x_*$ and $u_*$ is 
\begin{equation*}
\begin{split}
\VA[f_*|x_*, u_*, \omega_*, \dataset] & =  1 + \mbox{tr}\left[\M{Q} (\V{c}_{\X{x}*} \V{c}_{\X{x}*}^T) \odot \M{\tilde{C}}_{\X{zz}} \right] - \left[\tilde{\V{y}}^T (\V{c}_{\X{x}*} \odot \V{\tilde{c}}_{\X{z}*})\right]^2,
\end{split}       
\end{equation*}
where $\M{\tilde{C}}_{\X{zz}}$ is a $N \times N$ matrix with its $(i,j)$th element equal to 
\begin{equation*}
\begin{split}
\int c_z(z_i, z_*) c_z(z_j, z_*) & p(z_* |u_*, \omega_*, \dataset) dz_* \\
&= \int \exp\left\{ - \frac{(z_{i}-z_{*})^2 + (z_{j}-z_{*})^2}{2b_z^2} \right \}p(z_* |u_*, \omega_*, \dataset)  dz_* \\
& = (1+2t_*^2/b_z^2)^{-1/2} \exp\left\{-\frac{(\nu_* - \frac{z_i + z_j}{2})^2}{b_z^2 + 2t_*^2}\right\} \exp\left\{-\frac{(z_i - z_j)^2}{2b_z^2}\right\}. 
\end{split}
\end{equation*}

\subsection{Hyperparameter Estimation} \label{sec:hyper}
To achieve the posterior mean and variance in \eqref{eq:posexp}, a set of unknown hyperparameters are needed to be estimated. The parameters to be estimated are $\sigma^2$, the covariance parameters for $c$ and $k$, and the parameters of the two mean functions, $\rho(\omega)$ and $g(u)$. There are two sets of covariance parameters, one set for $c(\cdot, \cdot)$ and another set for $k(\cdot, \cdot)$. Let $\Theta_c$ and $\Theta_k$ denote the two sets respectively. For example, when $c$ is an exponential covariance given in \eqref{eq:cov_c}, $\Theta_c = \{b_z\} \cup \{b_l, l = 1,...,d-1\}$. Let $\alpha$ denote the parameter for $\rho$, and let $\V{\beta}$ denote a $Q \times 1$ parameter vector for $g$ with the $q$th element equal to $\beta_q$. 

A popular approach to estimate the hyperparamters of a GP model is the marginal likelihood maximization \citep{rasmussen2006gaussian}. We follow the approach for estimating the hyperparameters of the proposed model. The log marginal likelihood of the parameters given all observations is 
\begin{equation*}
\begin{split}
L(\alpha, \V{\beta}, \Theta_k, \Theta_c, \sigma^2| \V{y}, \V{z}) = \log p(\V{y} | \V{z}) p(\V{z}) \\
= \log p(\V{y} | \V{z}) + \log p(\V{z}).
\end{split}
\end{equation*}
where $p(\V{y} | \V{z})$ is the density of the conditional probability $P(\V{y}|\V{z})$, and $p(\V{z})$ is the density of the probability $P(\V{z})$. Based on \eqref{eq:joint_z}, the probability $P(\V{z})$ follows a normal distribution $\mathcal{N}(\V\rho + \V{g}, \M{K}_{\X{\omega\omega}})$. With the parametric models \eqref{eq:param}, $\V{\rho} = \rho \V{1}$, where $\V{1}$ is a $N \times 1$ vector of ones, and $\V{g} = \V{\Pi} \V{\beta}$, where $\V{\Pi}$ is a $N \times Q$ matrix with its $(i, q)$th element equal to $\pi_q(u_i)$,. The log density $\log p(\V{z})$ only depends on the choice of the parameters, $\alpha, \V{\beta}$ and $\Theta_k$, 
\begin{equation*}
\log p(\V{z}) = -\frac{N}{2} \log (2\pi) - \frac{1}{2}|\M{K}_{\X{\omega\omega}}| - \frac{1}{2}(\V{z} - \rho \V{1} - \V{\Pi} \V{\beta})^T \M{K}_{\X{\omega\omega}}^{-1} (\V{z} - \rho \V{1}- \V{\Pi} \V{\beta}).
\end{equation*}
Based on \eqref{eq:joint_y}, the conditional probability of $\V{y}$ conditioned on $\V{z}$ follows a normal distribution $\mathcal{N}(\V{0}, \sigma^2 \M{I} + \M{K}_{\X{\tilde{x}\tilde{x}}} )$, and its log density only depends on $\sigma^2$ and $\Theta_c$, 
\begin{equation*}
\log p(\V{y}|\V{z}) = -\frac{N}{2} \log (2\pi) - \frac{1}{2}| \sigma^2 \M{I} + \M{K}_{\X{\tilde{x}\tilde{x}}}| - \frac{1}{2}\V{y}^T ( \sigma^2 \M{I} + \M{K}_{\X{\tilde{x}\tilde{x}}})^{-1} \V{y}.
\end{equation*}  
Therefore, the maximum likelihood estimates for $\alpha, \V{\beta}$ and $\Theta_k$ can be achieved independently of achieving the maximum likelihood estimates of the other parameters. 

\subsection{Extensions to a General Case} \label{sec:ext}
The mathematical modeling and derivation have been so far shown for an experiment with only one main factor manipulated by other factors. In this section, we extend the result for a general case with multiple main factors manipulated. Consider a general case that involves $d$ main factors and a real response variable $y$. Among the $d$ factors, $m (\le d)$ factors are manipulated by other manipulating factors. Let $z^{(j)}$ denote the $i$th main factor manipulated by a $p_j$-dimensional manipulating factor $u^{(j)}$ for $j = 1,...,m$. The manipulation precision is limited, so $z^{(j)}$ achieved by applying $u^{(j)}$ could have bias $\rho_j$ and random variation $\delta_j$,
\begin{equation*}
z^{(j)} = g_j(u^{(j)}) + \delta_j(\omega^{(j)}),
\end{equation*} 
where both of the bias and random variation are independent of the manipulating factor $u^{(j)}$ but only depends on other process characteristics $\omega^{(j)}$. As previously treated, the random variation $\delta_j$ is assumed to be a zero-mean GP with covariance function $k^{(j)}$. Given all observations from past experiments, $\dataset_m = \{x_i, \{z_i^{(j)}\}, \{u_i^{(j)}\}, \{\omega_i^{(j)}\}, y_i \}$, the posterior distributions of $z^{(j)}$ at $u^{(j)} = u^{(j)}_*$ and $\omega^{(j)} = \omega^{(j)}_*$ for different $j$'s can be achieved independently in the form of
\begin{equation}\label{eq:pred-dist3}
z^{(j)} |u^{(j)}_*, \omega^{(j)}_*, \dataset_m \sim 
\mathcal{N}(\nu^{(j)}_*, t^{(j)}_*).
\end{equation}
We will skip writing the expression for the mean $\nu^{(j)}_*$ and variance $t^{(j)}_*$, because it is a simple repetition of the results \eqref{eq:pred-dist2}. 

The experimental outcome $y_i$ under the $i$th factor setting is assumed a noisy observation of an unknown function $f$ with noise $\epsilon_i$,
\begin{equation*}
y_i = f(x_i, z_i^{(1)}, \ldots, z_i^{(m)}) + \epsilon_i,
\end{equation*}  
where $f$ is a zero-mean GP with covariance function $c$. Assume that the covariance is defined as a product of the coordinate-wise covariance functions such as exponential or squared exponential covariances, 
\begin{equation}
c( (x_i, z_i^{(1)}, \ldots, z_i^{(m)}), (x_l, z_l^{(1)}, \ldots, z_l^{(m)})) = c_x(x_i, x_l) \prod_{j=1}^m c_{j}(z_i^{(j)}, z_l^{(j)}).
\end{equation}
Let $f_* = f(x_*, z_*^{(1)}, \ldots, z_*^{(m)})$ denote the value of $f$ at a test setting, $(x_*, z_*^{(1)}, \ldots, z_*^{(m)})$. The posterior distribution of $f_*$ given $x_*, z_*^{(1)}, \ldots, z_*^{(m)}$ is a multivariate normal distribution with mean $\V{\tilde{y}}^T \V{c}_{\X{\tilde{x}}*}$ and variance $c_{**} - \V{c}_{\X{\tilde{x}}*}^T (\sigma^2 \M{I} + \M{C}_{\X{\tilde{x}}\X{\tilde{x}}})^{-1}\V{c}_{\X{\tilde{x}}*}$. Based on this posterior distribution and the posterior \eqref{eq:pred-dist3}, the posterior expectation of $f_*$ given $x_*$ and $\{u^{(j)}_*, \omega^{(j)}_*\}$, is given in the form of 
\begin{equation*}
\V{\tilde{y}}^T (\V{c}_{\X{x}*} \odot \V{\tilde{c}}_{1*} \odot \cdots \odot \V{\tilde{c}}_{m*}), 
\end{equation*}
where $\V{\tilde{c}}_{j*}$ is a $N \times 1$ vector with the $i$th element equal to $\int c_j(z^{(j)}_i, z^{(j)}_*) p(z^{(j)}_* |u^{(j)}_*, \omega^{(j)}_*, \dataset_m) dz^{(j)}_*$. The value of the $i$th element should be given in the same form as Section \ref{sec:exp} for an exponential covariance and as Section \ref{sec:exp2} for a squared exponential covariance.  Similarly, the posterior variance of $f_*$ given $x_*$ and $\{u^{(j)}_*, \omega^{(j)}_*\}$, is given in the form of 
\begin{equation*}
\begin{split}
1 + \mbox{tr}\left[\M{Q} (\V{c}_{\X{x}*} \V{c}_{\X{x}*}^T) \odot \M{\tilde{C}}_{1} \odot \cdots \odot \M{\tilde{C}}_{m} \right] - \left[\V{\tilde{y}}^T (\V{c}_{\X{x}*} \odot \V{\tilde{c}}_{1*} \odot \cdots \odot \V{\tilde{c}}_{m*})\right]^2,
\end{split}       
\end{equation*}
where $\M{Q} = \tilde{\V{y}}\tilde{\V{y}}^T - (\sigma^2\M{I} + \M{C}_{\X{\tilde{x}\tilde{x}}})^{-1}$, and $\M{\tilde{C}}_{j}$ is a $N \times N$ matrix with its $(i,l)$th element equal to $\int c_j(z^{(j)}_i, z^{(j)}_*) c_j(z^{(j)}_l, z^{(j)}_*) p(z^{(j)}_* |u^{(j)}_*, \omega^{(j)}_*, \dataset_m) dz^{(j)}_*$.

\section{Simulation Study} \label{sec:sim}
This section presents our simulation study designed to show how the proposed approach works for various test scenarios. Primarily, we are interested in three major aspects of the proposed model, (1) the identifiability of $g(u)$ and $\delta(\omega)$, (2) the predictability of $z$, and (3) propagation of the errors from $z$ to $y$. Simulation data are generated with two main factors, $x$ and $z$, and $z$ is influenced by a manipulating factor $u$ and an exogenous variable $\omega$. Below is the description of the data generation procedure:

\begin{figure}
	\includegraphics[width=\textwidth]{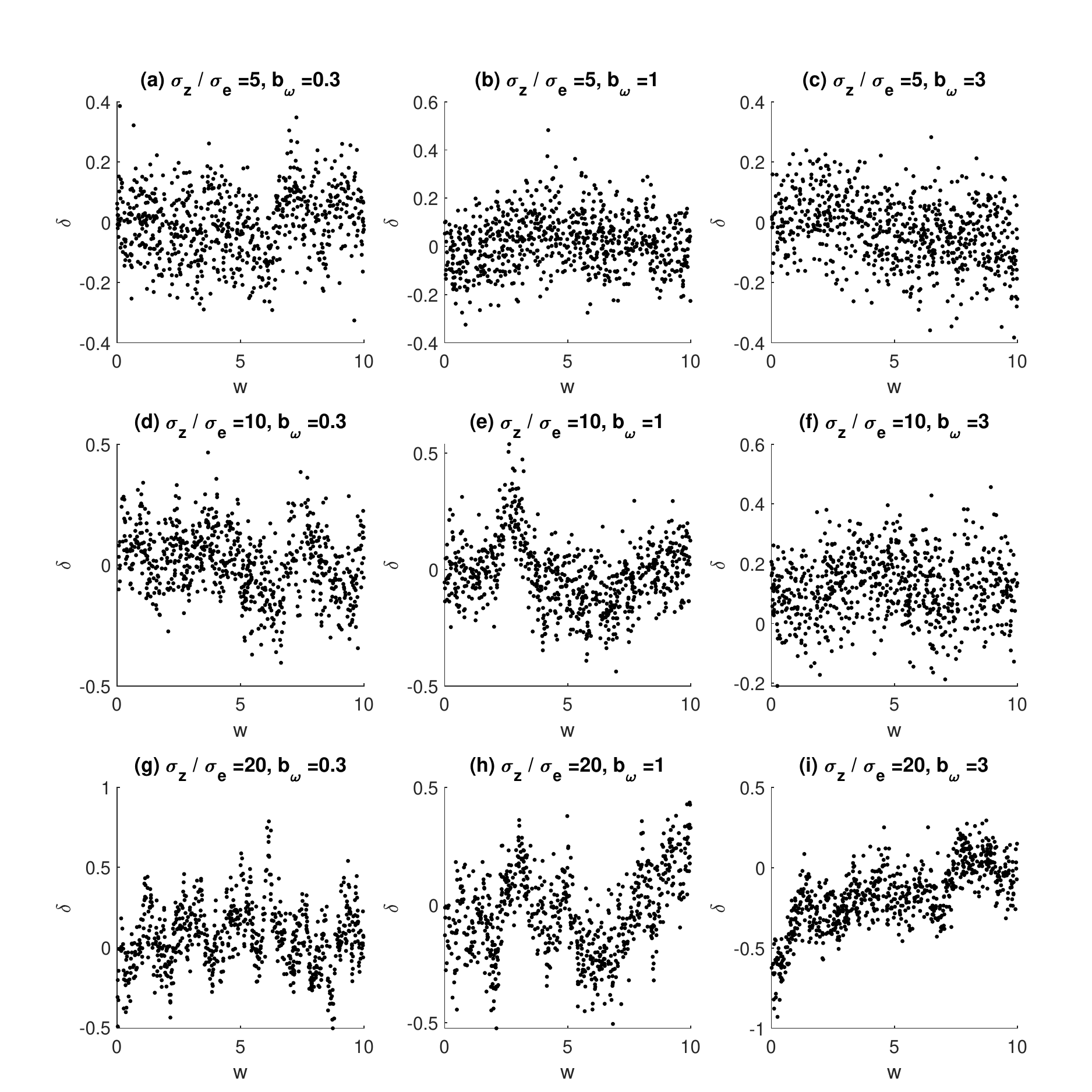}
	\caption{How $\delta(\omega)$ looks like for different simulated scenarios. } 
	\label{fig_sim_data}
\end{figure}

\begin{description}
	\item[\textbf{Step 1}.] Each of the independent variables, $u_i$, $x_i$ and $\omega_i$, is randomly sampled from the following uniform distribution,
	\begin{equation*}
	\Unif([0, 10]) \mbox{ for } i = 1,\ldots, N. 
	\end{equation*}  
	\item[\textbf{Step 2}.] The function $\delta$ is a realization of a zero mean Gaussian process with the exponential covariance function,  
	\begin{equation} 
	\begin{split}
	k(\omega_i, \omega_j) & = \sigma^2_z \exp\left\{ - \frac{|\omega_i-\omega_j|}{b_{\omega}} \right \},
	\end{split}
	\end{equation}
	and we used $g(u_i) = 1 + 0.5 u_i - 0.2 (u_i-5)^2$. Given the model, $\V{z}=(z_1, \ldots, z_N)^T$ is sampled from a multivariate normal distribution with mean $\V{g} = (g(u_1), g(u_2), \ldots, g(u_N))^T$ and covariance matrix $\M{K}$ whose $(i,j)$th element is $k(\omega_i, \omega_j)$. 
	\item[\textbf{Step 3}.] The main regression function $f$ is a realization of a zero mean Gaussian process with the exponential covariance function,
	\begin{equation} 
	\begin{split}
	c((x_i,z_i), (x_j, z_j)) & = \sigma^2_y \exp\left\{ - \frac{|x_{il}-x_{jl}|}{b_x} \right \} \exp\left\{ - \frac{|z_{il}-z_{jl}|}{b_z} \right \}.
	\end{split}
	\end{equation}
	This means $\V{y} = (y_1, \ldots, y_N)^T$ is sampled from a multivariate normal distribution with zero mean and covariance matrix whose $(i,j)$th element is $c((x_i,z_i), (x_j, z_j))$. 
	\item[\textbf{Step 4}.] 50 percent of the $N$ samples was randomly drawn to serve training data, and the remaining 50 percent serve test data. For the training samples, we added a Gaussian noise $\mathcal{N}(0, \sigma_e^2)$ to both of the response value $y_i$ and the main factor $z_i$. 
\end{description}

\subsection{First batch of simulation study} \label{sec:sim_first}
In the first batch of the simulation study, we mainly varied the two parameters, $b_{\omega}$ and $\sigma_z$, that affect the variation of $z$ due to the exogenous variable $\omega$, and we like to see how the estimation of $g$ is affected by the variation. The covariance range parameter $b_{\omega}$ varied over three possible values, 0.3 (short range), 1 (medium range), and 3 (long range). While fixing the noise variance parameter $\sigma_e=0.1$, we varied the signal variance parameter $\sigma_z$ over $0.5, 1$ and $2$, which makes the signal-to-noise ratio $\sigma_z/\sigma_e \in \{5, 10, 20\}$. The other parameters are fixed with $b_x = b_z = 1$, $\sigma_y / \sigma_e$ = 10, and $N = 1,500$.  In total, we have nine simulation scenarios, and we perform replicated experiments of 25 runs for each scenario. Without explicit description, all results presented in this section are averaged over the replications. Figure \ref{fig_sim_data} illustrates some representative trends of $\delta(\omega)$ that creates deviates from $g(u)$ for different values of $b_{\omega}$ and $\sigma^2_z$. 

In applying the proposed approach, the unknown function $g(u)$ are defined in the form of \eqref{eq:param} with polynomial basis functions,
\begin{equation*}
\begin{split}
\pi_q (u) = u^{q-1},
\end{split}
\end{equation*} 
and $Q = 5$. We used the exponential covariance function, and the hyperparameters are estimated as described in Section \ref{sec:hyper}. 

\begin{figure}
	\includegraphics[width=\textwidth]{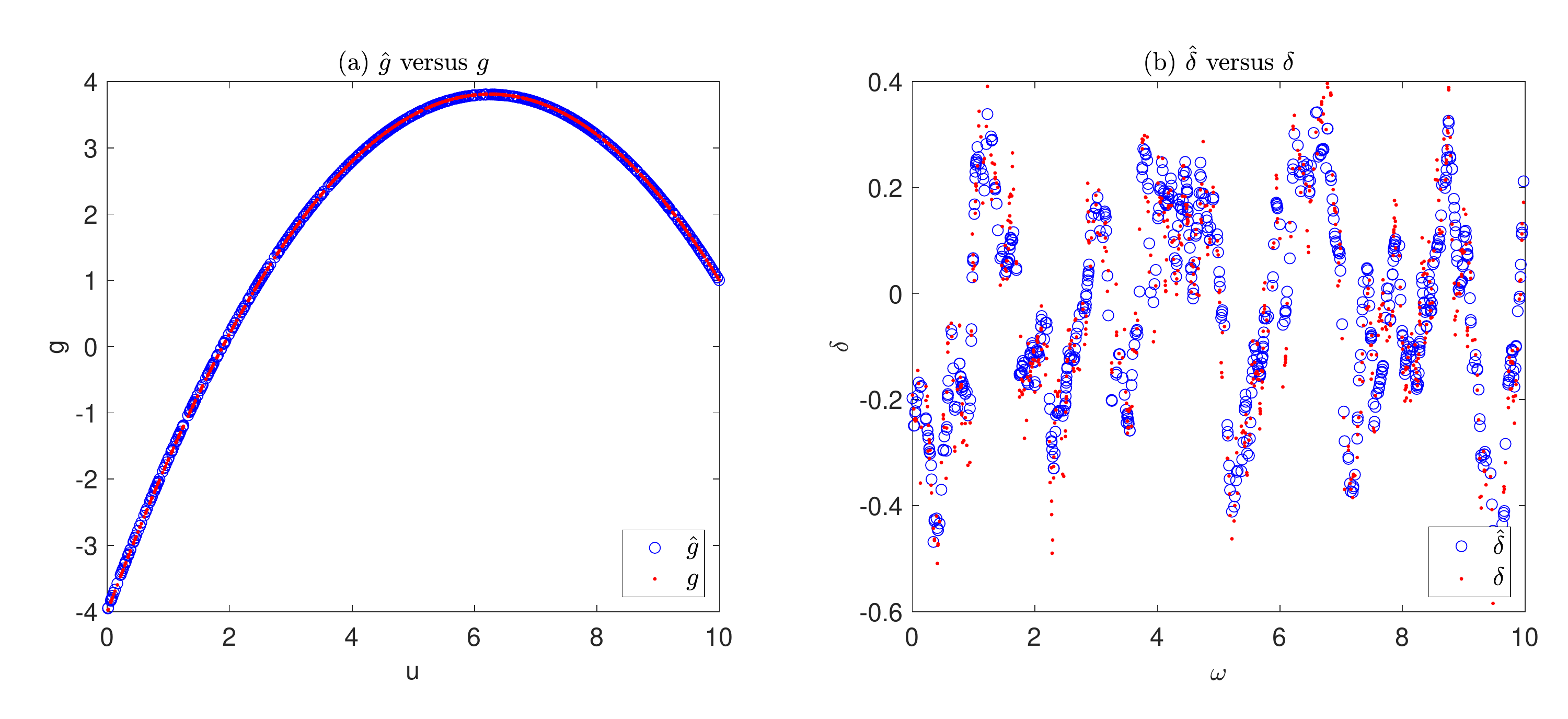}
	\caption{Estimation of $g(u)$ and $\delta(\omega)$ for $b_{\omega}=0.3$ and $\sigma_z / \sigma_e = 20$.} 
	\label{fig_sim_g}
\end{figure}

First, we looked at how the estimated $\hat{g}(u)$ is compared to $g(u) = 1 + 0.5 u_i - 0.2 (u_i-5)^2$ at the test locations to see how $g(u)$ is identifiable from the noisy observations of $z$. Table \ref{tbl_g} summarizes the mean absolute errors of $\hat{g}$ for different simulation scenarios. The estimation errors are not significant relative to the overall estimation error in $z$ (that will be reported in the next paragraph) for all of the test scenarios as shown in Figure \ref{fig_sim_g}-(a). This implies $g(u)$ is identifiable for various cases of $\delta$. 

We also looked at how the estimated $\hat{z}(u)$ is compared to $z$ for the test data. Table \ref{tbl_z} summarizes the mean absolute errors of $\hat{z}$ for different simulation scenarios. The errors of $\hat{z}$ are significantly higher than those of $\hat{g}$, which implies the major error source is the estimation of $\delta(\omega)$ as shown in Figure \ref{fig_sim_g}-(b). The error magnitudes were lower for larger covariance range parameter values and smaller ratios of $\sigma_z / \sigma_e$. For a large covariance range and a small ratio of $\sigma_z / \sigma_e$, $\delta(\omega)$ is estimable better, which gives a better accuracy in estimating $z$.

\begin{table}
	\centering
	\begin{tabular}{|c|c|c|c|}
		\hline
							 	& $b_{\omega}=0.3$   	& $b_{\omega}=1$  			& $b_{\omega} = 3$ \\ \hline
		$\sigma_z/\sigma_e = 5$ 	& 0.0044 (0.0021) 	&  0.0036 (0.0020) 		&  	0.0038 (0.0017)	       \\
		$\sigma_z/\sigma_e =10$ 	& 0.0049 (0.0026) 	&  0.0032 (0.0020) 		&  	0.0037 (0.0017)			\\
		$\sigma_z/\sigma_e =20$ 	& 0.0054 (0.0022)   &  0.0039 (0.0024) 		& 	0.0043 (0.0021)			\\ \hline
	\end{tabular}
	\caption{Mean absolute errors of the estimated $g$ against test data for different simulation scenarios. The results were averaged over 25 replications. The first number in each cell is the average, and the second number in round brackets is the standard deviation.}
	\label{tbl_g} 
\end{table}

\begin{table}
	\centering
	\begin{tabular}{|c|c|c|c|}
		\hline
								& $b_{\omega}=0.3$   	& $b_{\omega}=1$  			& $b_{\omega} = 3$ \\ \hline
		$\sigma_z/\sigma_e = 5$ 	& 0.0228 (0.0024) 	&  0.0167 (0.0016) 		&  	0.0132 (0.0019)	       \\
		$\sigma_z/\sigma_e =10$ 	& 0.0320 (0.0019) 	&  0.0236 (0.0016) 		&  	0.0179 (0.0015)			\\
		$\sigma_z/\sigma_e =20$ 	& 0.0471 (0.0023)   &  0.0342 (0.0017) 		& 	0.0257 (0.0015)			\\ \hline 
		\hline
	\end{tabular}
	\caption{Mean absolute errors of the estimated $z$ against test data for different simulation scenarios. The results were averaged over 25 replications. The first number in each cell is the average, and the second number in round brackets is the standard deviation.}
	\label{tbl_z} 
\end{table}

Last, we look at how the estimated $\hat{y}$ is compared to $y$ for the same simulation scenarios. The errors in $z$ vary from 0.0132 to 0.0471 under the scenarios, and we are primarily interested in seeing how the errors in $z$ are propagated to $y$. Table \ref{tbl_y} summarizes the mean absolute errors of $\hat{y}$ for different simulation scenarios. The errors in $y$ are not much affected by the errors in $z$, which may be mainly because we chose the covariance function for the $y$ process with mid covariance range and mid variance parameters. Under the chosen covariance, small errors in $z$ between 0.0132 and 0.0471 may not induce significant difference in the estimation of $y$. Therefore, we designed the second batch of simulation study as described in Section \ref{sec:sim_second}. 

\subsection{Second batch of simulation study} \label{sec:sim_second}
We performed a supplementary simulation study to see how the errors in $z$ are propagated to the errors in $y$ for different covariance models of $y$. In the supplementary study, we mainly tried different values of the covariance range and the variance parameter for the covariance function $c$. To be specific, we varied $b_z \in [0.3, 1, 3]$ and also varied $\sigma_y/\sigma_e$ over 5, 10 and 20 while fixing $b_{\omega}=1$ and $\sigma_z/\sigma_e = 10$. Please note that for $b_{\omega}=1$ and $\sigma_z/\sigma_e = 10$, the estimation error for $z$ is 0.0236 on average. We would like how this error is propagated to estimating $y$ when we have different covariance ranges and variance parameters for the response variable.  Table \ref{tbl_y2} shows the estimation error for $y$ with varying $b_z$ and $\sigma_y$. The overall error increases as $b_z$ decreases and $\sigma_y$ increases. If $b_z$ decreases, it would induce more local changes in $y$, so a small difference in $z$ would affect $y$ significantly. 

The major findings in the simulation study of Sections \ref{sec:sim_first} and \ref{sec:sim_second} are summarized as follows:
\begin{itemize}
	\item The effect $g(u)$ of the manipulating factor $u$ to the main factor $z$ can be identifiable for the additive model \eqref{eq:z} for various testing scenarios. 
	\item The accuracy of estimating $z$ is mainly limited by how $\delta(\omega)$ is estimable. The variance of $\delta$ and the covariance range of $\delta$ could affect the accuracy. Low variance and long covariance range parameters generally give better estimation accuracy for $\delta$ and so do for $z$.
	\item The error in estimating $z$ is propagated to the error in estimating the main response variable $y$. The propagation is mainly affected by the covariance structure of $y$, i.e., how $y$ changes under a small perturbation in $z$. When the covariance has a short range with respect to $z$, the error in $z$ is more amplified. In addition, the error is more amplified for a larger variance parameter $\sigma^2_y$. 
\end{itemize}

\begin{table}
	\centering
	\begin{tabular}{|c|c|c|c|}
		\hline
		& $b_{\omega}=0.3$   	& $b_{\omega}=1$  			& $b_{\omega} = 3$ \\ \hline
		$\sigma_z/\sigma_e = 5$ 	& 0.0526 (0.0049) 	&  0.0526 (0.0052) 		&  	0.0533 (0.0041)	       \\
		$\sigma_z/\sigma_e =10$ 	& 0.0525 (0.0037) 	&  0.0528 (0.0037) 		&  	0.0529 (0.0059)			\\
		$\sigma_z/\sigma_e =20$ 	& 0.0520 (0.0028)   &  0.0533 (0.0044) 		& 	0.0522 (0.0042)			\\ \hline
		\hline
	\end{tabular}
	\caption{Mean absolute errors of the estimated $y$ against test data for varying $b_{\omega}$ and $\sigma_z$, while fixing $b_z=1$ and $\sigma_f/\sigma_e=10$. The results were averaged over 25 replications. The first number in each cell is the average, and the second number in round brackets is the standard deviation.}
	\label{tbl_y} 
\end{table}

\begin{table}
	\centering
	\begin{tabular}{|c|c|c|c|}
		\hline
									& $b_{z}=0.3$   	& $b_{z}=1$  			& $b_{z} = 3$ \\ \hline
		$\sigma_y/\sigma_e = 5$ 	& 0.0372 (0.0016) 	&  0.0294 (0.0025) 		&  	0.0215 (0.0020)	       \\
		$\sigma_y/\sigma_e =10$ 	& 0.0688 (0.0040) 	&  0.0543 (0.0046) 		&  	0.0378 (0.0036)			\\
		$\sigma_y/\sigma_e =20$ 	& 0.1339 (0.0116)   &  0.0990 (0.0101) 		& 	0.0710 (0.0076)			\\ \hline
		\hline
	\end{tabular}
	\caption{Mean absolute errors of the estimated $y$ against test data for varying $b_{z}$ and $\sigma_y$. The results were averaged over 25 replications. The first number in each cell is the average, and the second number in round brackets is the standard deviation.}
	\label{tbl_y2} 
\end{table}

\section{GP Surrogate for Nanotube Growth Experiment} \label{sec:numerical}
We use our motivating example of a research robot, Autonomous Research System (ARES), that performs closed-loop carbon nanotube (CNT) growth experiments, to illustrate and validate the proposed method. For the past two decades, CNTs have been at the forefront of
nanotechnology \citep{rao2018carbon} but their use in large-scale applications has been hampered by our inability to produce them with controlled structures and properties. In the growth experiment, single wall carbon nanotubes are grown using a cold-wall chemical vapor decomposition reactor that is coupled to a Raman spectrometer, which provides the feedback for the closed-loop process. The detailed description of the experimental setup and the growth experiments can be found in our previous works \citep{nikolaev2016autonomy, rao2012situ, nikolaev2014discovery}. Within a single CNTs growth experiment, a growth condition is specified by seven factors, including a reaction temperature, a pressure, the flow rates of three gases (Ar, C$_2$H$_4$, H$_2$) and the concentration of water vapor. For each experiment, the growth rate of the CNTs is experimentally measured using \textit{in situ} Raman spectroscopy. Among the seven factors, the reaction temperature is controlled by applying laser heating, and the applied laser power is a manipulating factor to the temperature. Therefore, the experimental design of this growth experiment is defined by a design matrix of the manipulating factor and the other six main factors. Based on a prescribed experimental design, 719 experiments were performed \citep{nikolaev2016autonomy}. The major goal of studying this dataset is to build a predictive model of predicting the growth rate for a future experiment, and the predictive model will be used as a surrogate model to optimize the growth experiment in the Bayesian optimization. 

\begin{figure}
	\includegraphics[width=\textwidth]{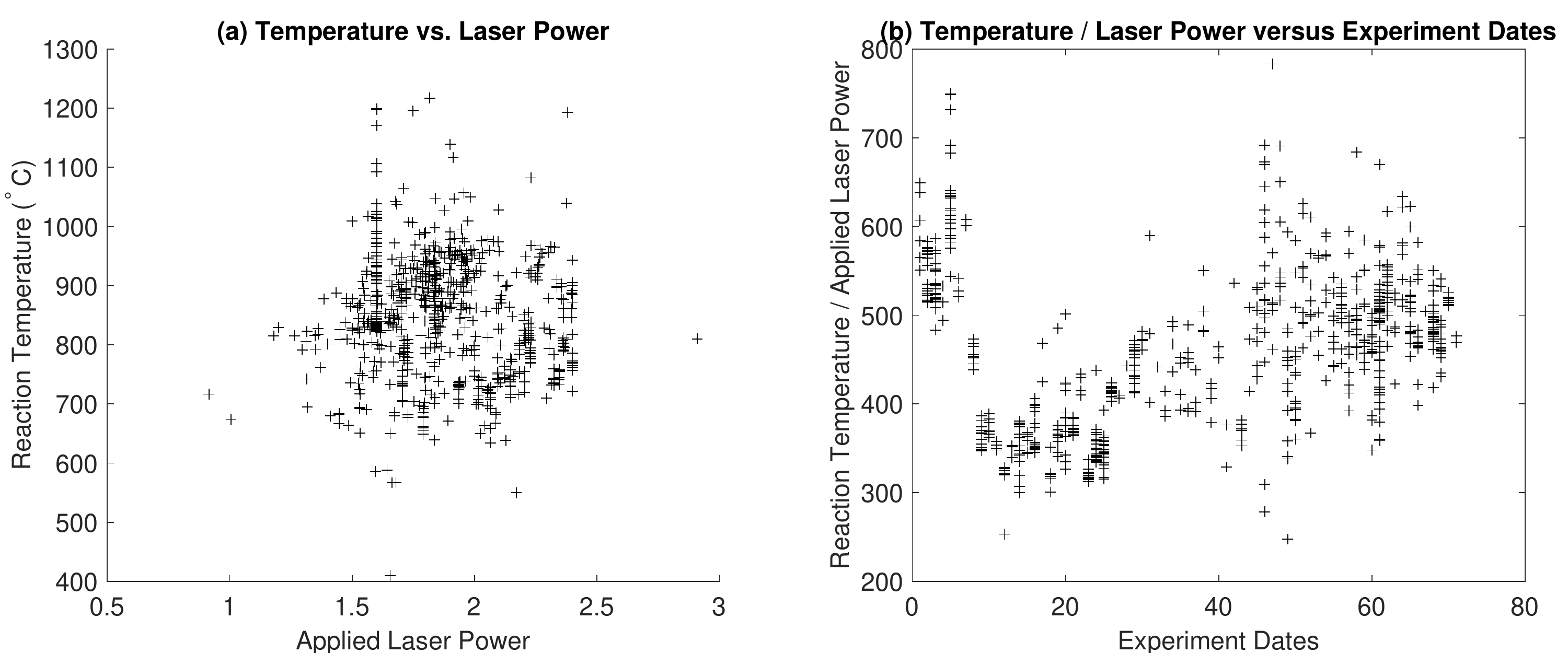}
	\caption{Control precision of the reaction temperature in nanotube growth experiment; (a) shows the relation of applied laser powers and the resulting temperatures, and (b) shows the variation of the relation for different experiment dates.} 
	\label{fig1}
\end{figure}

The major issue in achieving the goal is that the control precision to the reaction temperature is not high, so the reaction temperature achieved for a given laser power varies with a mean shift and a random variation. Figure \ref{fig1}-(a) shows the experimentally achieved temperatures for applied laser powers from the 719 experiments. From the first look of the scatter plot, the applied laser powers and the resulting temperatures appear mutually independent. However, if we split the experimental outcomes by the dates of the experiments and plot the ratio of the temperature and the applied laser power over different dates of the experiments as shown in Figure \ref{fig1}-(b), the correlation between the laser power and the temperature is revealed more clearly. In the figure, the ratio values are concentrated around one center per each date of the experiments, which implies that the laser power and temperature have some correlation. However, the center value changes in the dates, so aggregating all of the data with no consideration of the dates of the experiments would blur this correlation. Therefore, the straightforward application of the Gaussian process model to relate the laser power and other six factors to the growth rate would not provide a great predictive power of the growth rate as shown in Table \ref{tbl1}. On the other hand, a Gaussian process model that relates the temperature and the six factors to the growth rate does not help in predicting a future experiment, because the temperature for the future experiment is only observed after the experiment is done.  

To build a better predictive model, we applied our two-tiered GP modeling approach to relate the applied laser power and other main factors to the nanotube growth rate. In our model, the laser power becomes the manipulating factor $u$, the date of experiment is $\omega$, and the reaction temperature is the manipulated main factor $z$, while the remaining six main factors are composing $x$. The first tier in our two-tiered GP model relates $u$ to $z$ through an unknown function $g$ under a source of variation $\omega$, and the 
second tier relates $z$ and $x$ to the growth rate $y$ through another unknown function $f$. 

\subsection{Tier-1 GP model \eqref{eq:z} for temperature vs. laser} 
The relation of the applied laser power $u$ to the resulting temperature $z$ varies from day-to-day mainly due to variations in laser alignment, which affects the focused spot area. Therefore, the experiment operation dates define an additional process characteristic $\omega$ that affects $z$, so the relation of $u$ and $z$ is dependent on $\omega$ as shown in Figure \ref{fig1}-(b), where the ratios of the observed temperature $z$ and the applied laser power $u$ are plotted for different experimental dates $\omega$. The ratios are centered differently, and the variation of the ratios change from days to days. Certainly, the $u$-to-$z$ relation could be more complicated than the linear proportionality but this plot of the ratios is just shown to illustrate the dependency of the relation on $\omega$. 

We applied our the modeling \eqref{eq:z} to model the dependency of laser power on temperature. Specifically, $\delta$ is modeled as a realization of a zero mean GP with $k$ as an exponential covariance function. The relation of $u$ to $z$ should be monotonic, because a higher laser power results in a higher temperature. The function $g(u)$ is defined in the form of \eqref{eq:param} with polynomial basis functions $\pi_q$,
\begin{equation*}
\begin{split}
\pi_q (u) = u^{q-1}.
\end{split}
\end{equation*} 
The hyperparameters of the covariance function $k$ and the parameters of $\rho$ and $g$ were learned using the marginal likelihood maximization described in Section \ref{sec:hyper}. The posterior predictive distribution of the temperature at a test input $u_*$ and $\omega_*$ was achieved using the result \eqref{eq:pred-dist2}. 

\begin{figure}[ht!]
	\includegraphics[width=\textwidth]{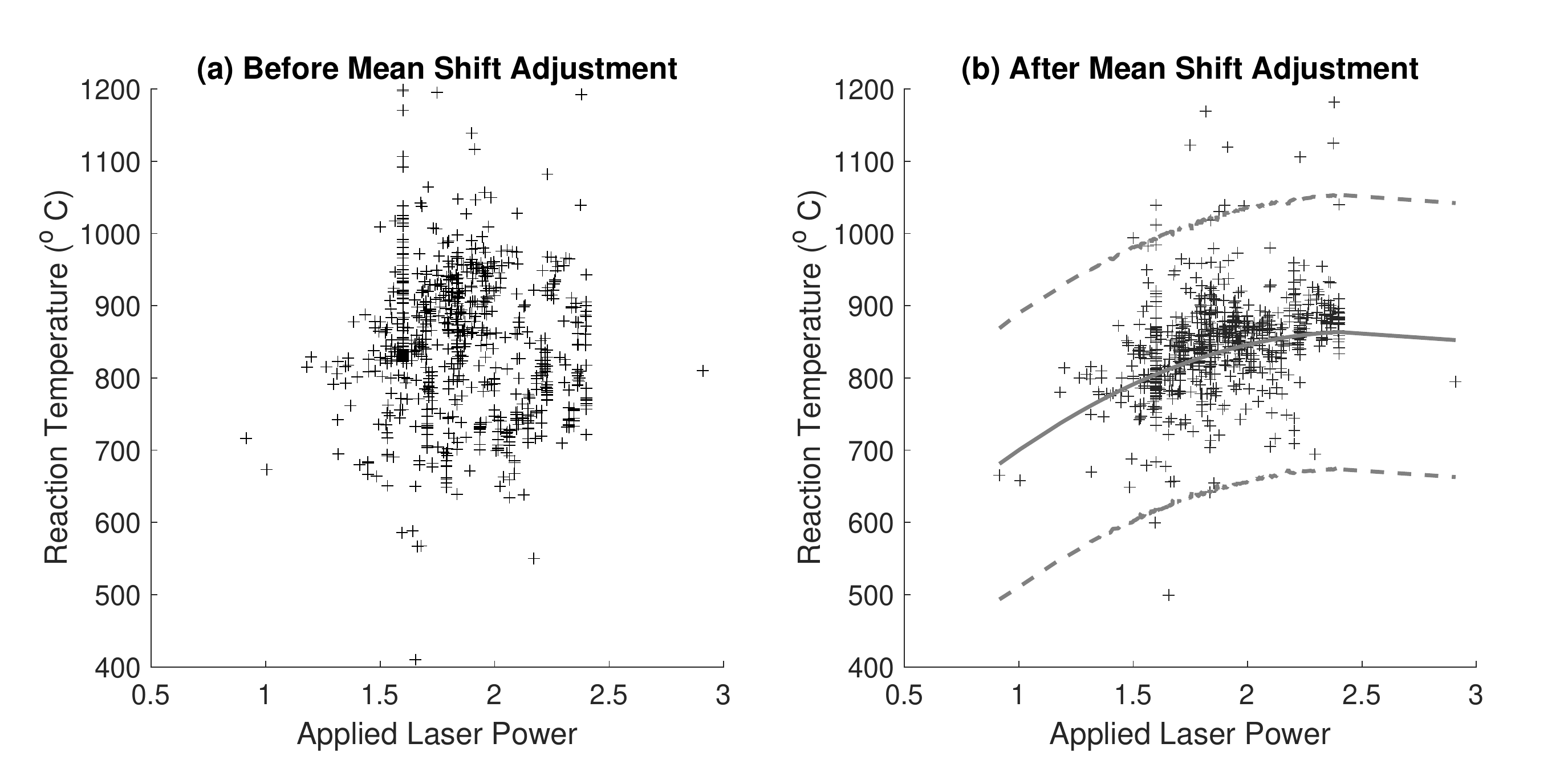}
	\caption{Temperature versus Laser Power. (a) Raw Temperature versus Laser Power, (b) Temperature minus estimated bias $\rho$ versus Laser Power. In (b), the center line is the mean prediction, while the two dotted lines represent the predicted mean $\pm$ 1.5 times of the predicted variance. } 
	\label{fig2}
\end{figure}

Considering the estimated mean shift in $z$ due to $\omega$, we first looked at how the laser power $u$ is related to the reaction temperature $z$. Figure \ref{fig2}-(b) shows the relation after the estimated posterior mean of $\delta(\omega)$ is subtracted from $z$, which is compared with the raw data in Figure \ref{fig2}-(a). After the mean subtraction, a much clearer relation is revealed. This also improved the prediction of $z$ given $u$. We randomly split 719 experimental records into a training data set and a test dataset. The training dataset contains 60\% of the 719 experimental records, while the test dataset contains the remaining 40\%. The proposed GP model \eqref{eq:z} was fitted using the training data set, and the fitted model was used to predict $z$ for the experimental settings in the test data set. The mean squared prediction error was $3.703 \times 10^3$, which is much lower than $7.1300 \times 10^3$, the prediction error of a standard GP model with $u$ as an input and $z$ as an output. 

\subsection{Combined GP metamodeling for predicting the nanotube growth rates}
\begin{figure}[ht!]
	\includegraphics[width=\textwidth]{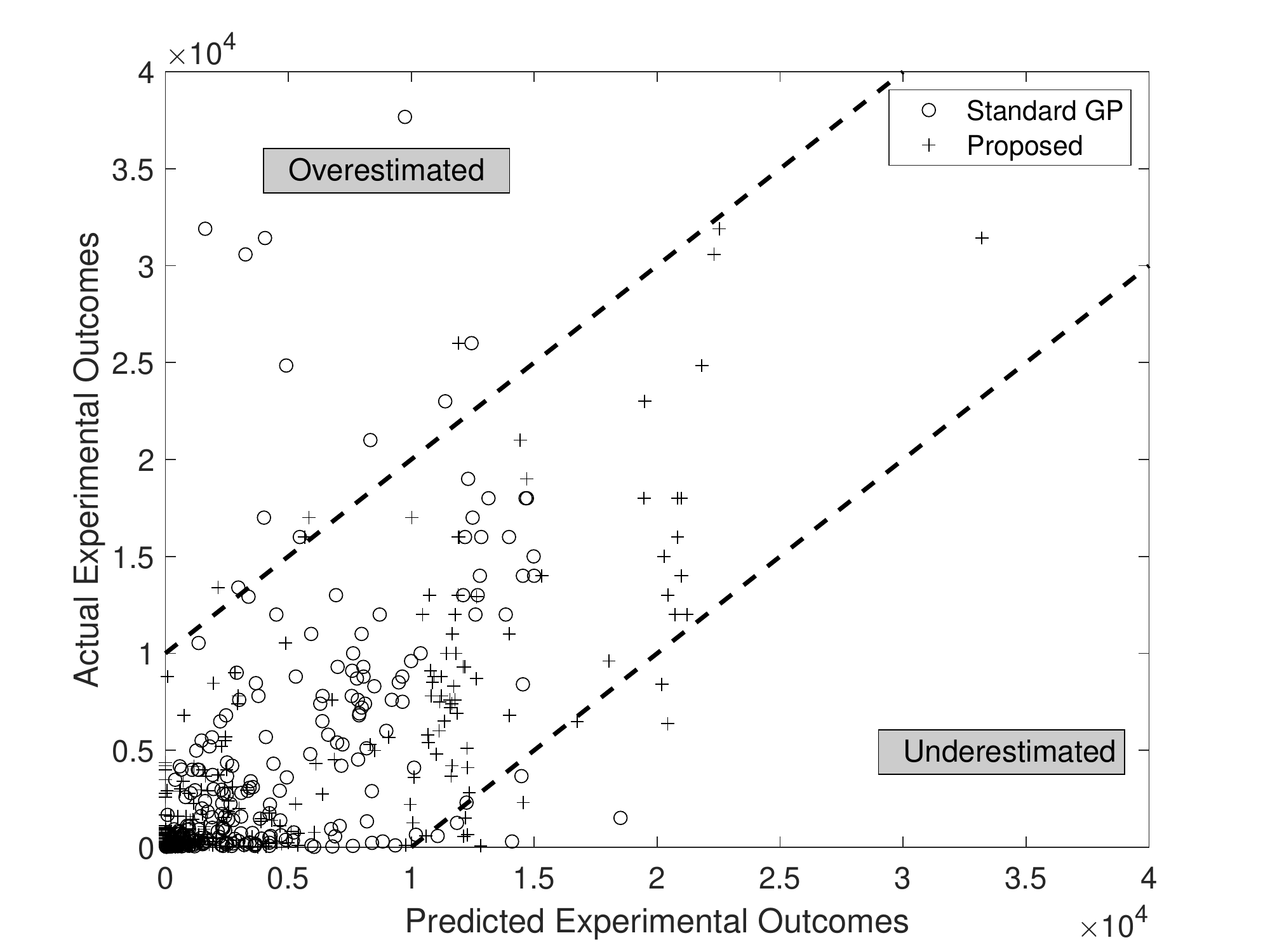}
	\caption{Predicted Outcomes Versus Actual Experimental Outcomes. The two dotted lines are added to distinguish `good' estimates (points in between the two dotted lines) versus the under-estimated (points above the left dotted line) versus the over-estimated (points below the right dotted line).} 
	\label{fig3}
\end{figure}

The relation $f$ relating $x$ and $z$ to the growth rate $y$ is assumed a realization of a zero-mean GP with an exponential covariance function. The hyperparameters of the covariance function $f$ were learned using the marginal likelihood maximization described in Section \ref{sec:hyper}. The model was combined with the Tier-1 GP model described in the previous section to achieve the posterior predictive distribution of $f$ at a test setting of $u_*$ and $z_*$. For evaluating the prediction accuracy, we randomly split 719 experimental records into a training data set and a test dataset. The training dataset contains 60\% of the 719 experimental records, while the test dataset contains the remaining 40\%. The proposed GP model was fitted using the training data set, and the fitted model was used to predict the growth rates for the experimental designs in the test data set, denoted by $\{(x_t, u_t, y_t): t=1,\dots,T\}$, where $T$ is the size
of the test set. Let $\mu_t$ and $\sigma^2_t$ denote the estimated posterior predictive mean and variance at the $t$th test condition. The predicted values were compared using the test data set in terms of two performance measures. The first measure is the mean squared error (MSE)
\begin{equation}
\textrm{MSE} = \frac{1}{T} \sum_{t=1}^T (y_t - \mu_t)^2,
\end{equation}
which measures the accuracy of the mean prediction $\mu_t$ at location $x_t$.
The second measure is the negative log predictive density (NLPD)
\begin{equation}
\textrm{NLPD} = \frac{1}{T} \sum_{t=1}^T\left[ \frac{(y_t - \mu_t)^2}{2\sigma_t^2} + \frac{1}{2} \log (2\pi \sigma_t^2) \right].
\end{equation}
The NLPD quantifies the degree of fitness of the estimated predictive distribution $\mathcal{N}(\mu_t, \sigma_t^2)$ for the test data. The NLPD is not only affected by the posterior mean prediction but also affected by the posterior variance, so it can be used to evaluate how the model uncertainty estimated is fit to the test data. These two criteria are used broadly in the GP regression literature. A smaller value of MSE or NLPD indicates better performance.  As a benchmark, a standard GP model was also fitted, and its MSE and NLPD were computed. Table \ref{tbl1} summarizes the MSE and NLPD measures for the proposed approach and the standard GP approach, which are the numbers averaged over 25 replicated experiments of different random splits of the training and test datasets. Figure \ref{fig3} also compares the predicted outcomes versus the actual experimental outcomes. The standard GP model yielded a significant number of underestimations and overestimations of the growth rates, and the degrees of the wrong estimates are significant. As a result, the MSE and NLPD values of the standard GP model were much higher than those of the proposed method. The higher predictive power of the proposed method could be used to predict the future experiment outcome at a specified experimental setting, and it can be exploited to find the optimal experimental set-up that achieves the desirable experiment outcome. 

\begin{table}
	\centering
	\begin{tabular}{|c|c|c|}
		\hline
		& Proposed   & Standard GP \\ \hline
		MSE ($\times 10^7$) & 3.16 (0.81) &  4.94 (1.62) \\
		NLPD & 10.10 (0.0572) & 10.57 (0.2218) \\ \hline
	\end{tabular}
	\caption{MSE and NLPD Performance of the Proposed Approach and the Standard GP Approach. 25 replicated experiments were performed for different random splits of training and test datasets, and the averages and standard deviations of the results were reported in the table. The first number in each cell is the average, and the number in round brackets is the standard deviation.}
	\label{tbl1} 
\end{table}

\section{Conclusion and Practical Implication} \label{sec:conclusion}
We presented a new Gaussian process (GP) surrogate modeling approach for predicting the outcome of a physical experiment for a given input factor setting where some of the input factors are controlled by other manipulating factors. We specifically considered the case that the purpose of the GP surrogate is to predict and optimize the future experimental outcomes. For such cases, the experimental design would include the settings of the manipulating factors in its design matrix. Therefore, the corresponding surrogate model for the experiment should better take the manipulating factors as input variables and predicts the experimental outcome, because the main input factors cannot be directly set. However, the GP surrogate with the manipulating factors is very tricky when the manipulation precision of the input factors is limited, so the input factor values vary significantly even under the same setting of the corresponding manipulating factors. Due to this variation, the standard Gaussian process surrogate that directly relates the manipulating factors to the experimental outcome would not provide a great predictive power on the outcome. To address this challenge, we proposed a two-tiered Gaussian process modeling, where the bottom tier relates the manipulating factors to the corresponding main factors with potential biases and variation, and the top tier relates the main factors to the experimental outcome. The two-tier model explicitly models the propagation of the manipulation  uncertainty to the experimental outcome through the two GP modeling tiers. 

The proposed approach was applied to our motivating example of carbon nanotube growth experiment, providing a superior accuracy of predicting the carbon nanotube growth rates given a setting of a  manipulating factor and other main input factors. This excellent prediction power is practically useful to supplement expensive nanotube growth experiments with a cheap statistical inference step, and the surrogate modeling can be also used as an objective function to optimize for optimizing the growth experimental conditions to achieve the desirable growth rate and ultimately to improve the rate of convergence for ARES. 

\section*{Acknowledgment} 
We acknowledge support for this work from the AFOSR (FA9550-18-1-0144) and the prime contract of the U.S. Federal Government, Contract No. FA8650-15-D-5405. 

\bibliographystyle{imsart-nameyear} 
\bibliography{gpv2}

\end{document}